\documentclass{kapedbk} % Computer Modern font calls

\usepackage[dvips]{graphicx}

\chaptersection % will use chapter.section form for section heads.

\upperandlowercase

 \let\footnote\savefootnote

\kluwerbib % will produce this kind of bibliography entry:

\begin{document}

\articletitle[Compton Thick AGN]
{Compton Thick AGN:\\
The dark side of the X-ray\\
background}
\rhead{The dark side of the XRB}

\author{Andrea Comastri}

%% affil, email, and abstract are optional
\affil{INAF--Osservatorio Astronomico di Bologna\\
via Ranzani 1, I--40127, Bologna, Italy
\footnote{Partially supported by the Italian Space Agency
(ASI) under grants I/R/073/01 and I/R/057/02, by the MIUR grant Cofin--03--02--23 and by the INAF grant 270/2003}}
\email{comastri@bo.astro.it}

\begin{abstract}

The spectrum of the hard X-ray background records the history 
of accretion processes integrated over the cosmic time.
Several pieces of observational and theoretical evidence 
indicate that a significant fraction of the energy density 
is obscured by large columns of gas and dust.
The absorbing matter is often very thick, with column densities
exceeding $N_H\simeq 1.5\times 10^{24}$~cm$^{-2}$, the value
corresponding to unity optical depth for Compton scattering.
These sources are called ``Compton thick'' and appear 
to be very numerous, at least in the nearby universe.
Although Compton thick Active Galactic Nuclei (AGN) are thought 
to provide an important contribution to the overall cosmic energy 
budget, their space density and cosmological evolution are poorly 
known. The properties of Compton thick AGN are reviewed here, with 
particular emphasis on their contributions to the extragalactic 
background light in the hard X-ray and infrared bands.

\end{abstract}

%\begin{keywords}
%Active Galactic Nuclei -- X-ray -- Extragalactic Background Light
%\end{keywords}

\section{Introduction}

Most of the active galactic nuclei (AGN) in the local universe
are obscured in the X-ray band by large amounts of gas and dust,
which prevent the observation of their nuclear emission 
up to energies that depend on the amount of intrinsic absorption. 
If the X-ray obscuring matter has a column density which is equal 
to or larger than the inverse of the Thomson 
cross-section ($N_H\ge \sigma_T^{-1} \simeq 1.5 \times 10^{24}$~cm$^{-2}$), 
then the source is called, by definition, ``Compton thick''. 
The cross-sections for Compton scattering and photoelectric absorption 
have approximately the same value for energies of order 10~keV, 
which can be considered as the low energy threshold for probing the 
Compton thick absorption regime.
Indeed, if the column density does not exceed a value of order 
10$^{25}$~cm$^{-2}$, then the nuclear radiation is visible above 
10~keV, and the source is called mildly Compton thick. 
For higher column densities (heavily Compton thick), 
the entire high energy spectrum is down-scattered by Compton recoil
and hence depressed over the entire X-ray energy range. 
The presence of Compton thick matter may be inferred through indirect
arguments, such as the presence of a strong iron K$\alpha$ line 
complex at $6.4-7$~keV and the characteristic reflection spectrum. 

The study of Compton thick sources is relevant for several reasons:
(1) there is observational evidence that
suggests that a large fraction of AGN in the local universe
are obscured by Compton thick gas (Maiolino et al.\ 1998; 
Risaliti, Maiolino, \& Salvati 1999a; Matt et al.\ 2000); 
(2) a sizable population of mildly Compton thick sources is 
postulated in all the AGN synthesis models for the X-ray background 
(XRB) in order to match the intensity peak of the XRB spectrum 
at about 30~keV. The absorbed luminosity eventually will be reemitted 
in the far-infrared (far-IR), making Compton thick sources potential 
contributors to the long wavelength background. Finally, accretion
in the Compton thick AGN may contribute to the local black hole 
mass density.

Unfortunately, the most efficient energy range to search for mildly
Compton thick sources is just above the highest energy accessible 
to the past and present generation of satellites with
imaging capabilities for faint limiting fluxes.
As a consequence, the search for Compton thick sources has been 
limited, so far, to the relatively bright fluxes accessible to 
the high energy detectors onboard {\em BeppoSAX} and {\em RXTE}.

In this chapter, we review the evidence for obscured AGN, with
a special emphasis on the evidence for Compton thick sources. 
We then examine in some detail the contributions of Compton thick 
sources to the XRB.

\section{Absorption Distribution in the Local Universe}
\label{comastriseclocal}

According to the so-called AGN unified model (Antonucci 1993),  
Seyfert II galaxies are powered by the same engine 
(a supermassive black hole plus an accretion disk) as
Seyfert I galaxies, but they are viewed through a geometrically 
and optically thick structure of gas and dust with an axisymmetric 
geometry (known as the ``torus'') that absorbs the nuclear 
radiation at ultraviolet (UV)/optical and soft X-ray energies.  
The distinction between type I and type II galaxies 
originally was conceived to classify objects characterized 
by different properties at UV and optical wavelengths.
Both broad and narrow lines are visible over a strong 
blue UV-optical continuum in type 1 galaxies, while only 
narrow lines are observed in type 2 galaxies.

In the simplest version of the unified models, X-ray obscured 
sources are expected to be uniquely associated with optical 
type II sources. The increasing evidence of a mismatch between 
optical and X-ray classification (which will be discussed 
in the following sections) suggests that the terms type I and 
type II should be treated with caution and should always
be referenced to a specific band. Following Matt (2002), in this 
review, I will use the original meaning of type II and type I, 
which is based on optical spectroscopy. 

The amount of obscuring material can be measured best
in the hard ($2-10$~keV) X-ray band, which is transparent up to 
column densities of order 10$^{24}$~cm$^{-2}$ (i.e., Compton thin); 
hence, hard X-ray surveys are a powerful method to obtain large, 
unbiased samples of obscured AGN. Since the first observations, 
it has been realized that the large majority of Seyfert II galaxies 
contain obscured AGN, which are fairly strong X-ray sources with 
$2-10$~keV luminosities up to 10$^{44}$~ergs~s$^{-1}$.

\begin{figure}[ht]
\vskip -0.25cm
\centerline{\includegraphics[width=\textwidth, height=4in]{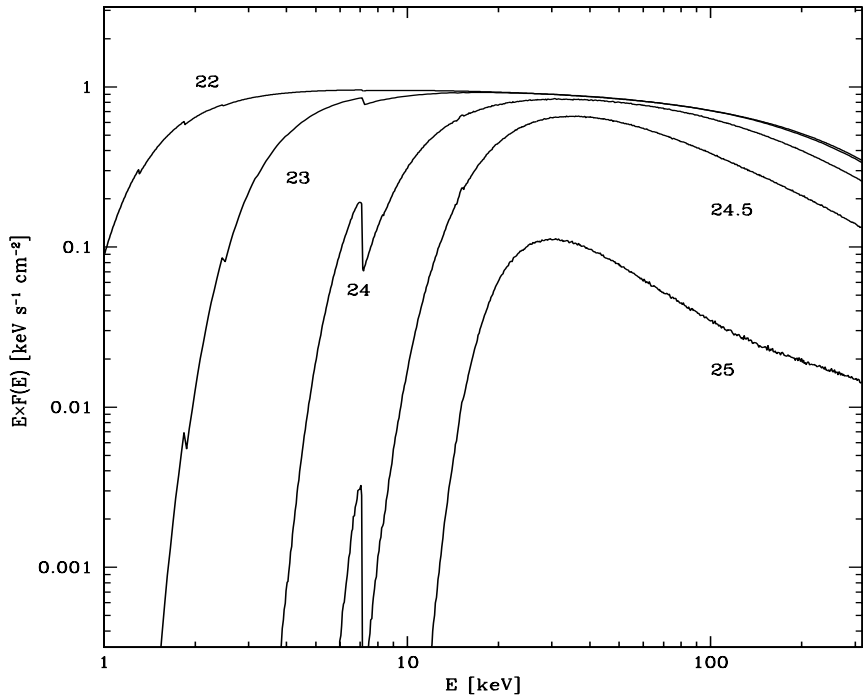}}
\vskip -0.5cm
\caption{Effects of photoelectric absorption and Compton 
down-scattering on the typical AGN X-ray spectrum (a power law 
with photon index $\Gamma=2$, plus an exponential cut-off at 300~keV). 
Labels correspond to the logarithm of the column density.}
\label{comastrifig1}
\end{figure}

The relative fraction of sources with a given column density 
depends on the survey sensitivity at high energies and on 
the amount of obscuring gas. The larger the column density, and/or 
the lower the sensitivity, the stronger will be the observational 
bias against the discovery and identification of obscured AGN.
Such a bias is maximal for (mildly) Compton thick sources due to 
the rapidly decreasing effective area at high energies of the 
most sensitive space observatories (the {\em Chandra} and 
{\em XMM-Newton X-ray Observatories}). Thus, we cannot measure 
the photoelectric cut-off and the corresponding $N_H$ value.
Moreover, even observations in the hard X-ray domain above 
10~keV do not allow the sampling of column densities of order
10$^{25}$~cm$^{-2}$ or larger because the entire high energy 
spectrum is down-scattered by Compton recoil to lower energies, 
where it is readily absorbed (see Fig.~\ref{comastrifig1}). 
The high energy spectra of obscured AGN are often much more 
complex than a single absorbed power law. Additional components 
(i.e., thermal emission from a starbursting region and/or nuclear 
flux scattered into the line of sight) are common in the X-ray 
spectra below 10~keV of several nearby Compton thick 
galaxies (Matt et al.\ 1997; Vignati et al.\ 1999). 
The photoelectric cut-off (if any) does not provide a measure of 
the ``true'' nuclear absorption anymore.

%
% FIGURE 1
%

A powerful diagnostic of the presence of obscuring, possibly 
Compton thick, matter is provided by the intensity of the iron 
line, which is expected to be produced both by transmission through
(Leahy \& Creighton 1993) and/or reflection by absorbing 
gas (Ghisellini, Haardt, \& Matt 1994; Matt, Brandt, \& Fabian 1996). 

As far as the transmitted continuum is concerned, 
the line equivalent width (EW) increases with the column density 
(since it is measured against an absorbed continuum) and reaches 
values of order 1~keV for $N_H\sim$~10$^{24}$~cm$^{-2}$. Larger 
values for the EW (up to several keV) can be obtained for high 
inclination angles and small torus opening angles 
(Levenson et al.\ 2002).

The signature of Compton thick matter is also imprinted 
on the ``reflected'' light, the so-called ``Compton reflection'' 
continuum characterized by a broad bump peaking around $20-30$~keV, 
which rapidly decreases at both low and high energies due to 
absorption and Compton down-scattering, respectively.
In the $2-10$~keV range, it is well approximated by a flat power law.
The iron line EW with respect to the reflected continuum is always 
greater than 1~keV (George \& Fabian 1991; Matt, Perola, \& Piro 1991).   

%
% FIGURE 2
%
\begin{figure}[h]
\centerline{\includegraphics[width=3.5in, angle=-90]{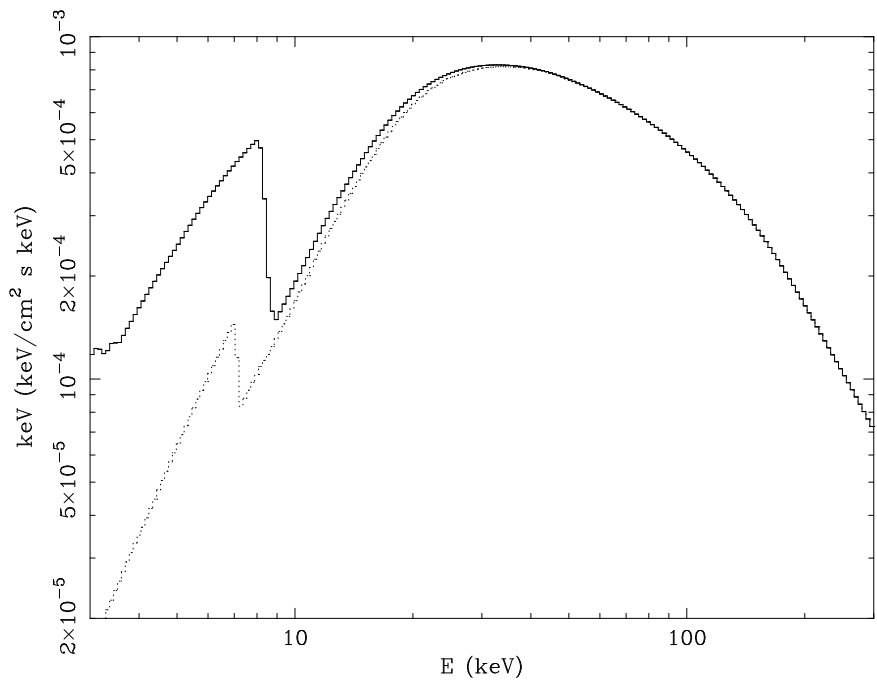}}
\caption{Spectrum reflected by Compton thick gas. Dotted curve 
corresponds to neutral gas; solid curve corresponds to highly 
ionized material.} 
\label{comastrifig2}
\end{figure}

The shape of the reflection spectrum (reported in 
Fig.~\ref{comastrifig2} for two specific examples)
is also a function of the column density, the ionization status, 
and the metallicity of the reflecting matter (Matt 2002). Its 
intensity depends upon the solid angle subtended by the reflector 
at the primary illuminating source. The gas is more reflective 
if highly ionized, and the effect is significant at energies 
below $7-8$~keV.  

Despite the presence of other spectral components, 
such as thermal emission from a hot plasma with a range 
of temperatures, and/or scattered nuclear light 
that may dilute the typical reflection spectrum,  
the signature of Compton thick matter (i.e., a strong iron 
K$\alpha$ line) has been unambiguously revealed already with 
the {\em ASCA} satellite in several nearby bright objects 
(i.e., the Circinus galaxy, Matt et al.\ 1996; 
NGC~6240, Iwasawa \& Comastri 1998).

Within the framework of the unified scheme, a further method 
for evaluating $N_H$ has been proposed by Bassani et al.\ (1999).
For a heterogeneous but sizable sample of Seyfert II galaxies, 
they found that the ratio between the $2-10$~keV and the dereddened 
[OIII] flux (assumed to be an isotropic indicator of the 
intrinsic luminosity) is anti-correlated with the absorption column 
density and the EW of the iron line. Such a relationship could be 
used to select candidate Compton thick AGN, especially among those  
weak sources for which X-ray spectroscopy is not feasible. 
In this regard, it is interesting to note that the Seyfert II 
galaxy NGC~5135, characterized by a very low value of 
$F_X/F_{\rm [OIII]}$ and suspected to be Compton thick, 
is now unambiguously classified as such, thanks to the 
detection of a strong iron line in the {\em Chandra} spectrum 
(Levenson et al.\ 2004).

A significant step forward in the study of highly obscured 
AGN has been made with the Phoswich Detector System (PDS) instrument 
onboard {\em BeppoSAX}, which made accessible, for the first time, 
the $10-100$~keV energy range down to limiting fluxes of order
10$^{-11}$~ergs~cm$^{-2}$~s$^{-1}$. Deep exposures of a sample 
of seven nearby, bright objects selected by the 
presence of the characteristic features described above
allowed the intrinsic nuclear spectrum above 10~keV
to be unambiguously uncovered in 5 out of the 7 sources, and 
column densities in the range $1-5\times 10^{24}$~cm$^{-2}$
to be measured (see Matt et al.\ 2000 for a review).

The issue of how common Compton thick sources are
and whether they constitute a sizable fraction of the Seyfert 
population has been addressed by Maiolino et al.\ (1998) and 
Risaliti et al.\ (1999a). Starting from a sample of local Seyfert II 
galaxies selected on the basis of their [OIII] flux, and using 
{\em ASCA} and {\em BeppoSAX} observations, they concluded that 
about half of the objects are obscured by column densities 
$N_H>10^{24}$~cm$^{-2}$. Given the lack of complete spectral 
coverage at energies $>10$~keV with {\em BeppoSAX}, the relative 
fraction of heavily ($N_H>10^{25}$~cm$^{-2}$) and mildly 
($10^{24}<N_H<10^{25}$~cm$^{-2}$) Compton thick sources remained 
poorly constrained.

Several independent arguments suggest that the space density of 
Compton thick sources, at least in the local universe, could be high.
For example, two out of the three nearest AGN within 4~Mpc 
are mildly Compton thick (NGC~4945 and Circinus; the third source, 
Cen A, is also obscured with $N_H\simeq 10^{23}$~cm$^{-2}$).
A simple estimate obtained by integrating the AGN luminosity 
function indicates that heavily obscured AGN could outnumber 
unobscured AGN by about one order of magnitude 
(see Matt et al.\ 2000 for a detailed discussion).  

The optical appearance of Compton thick AGN may also contribute
to raising the estimate of their space density. Indeed, two bright 
objects (NGC~4945 and NGC~6240) are classified as LINERs 
(Low Ionization Nuclear Emission Regions) on the 
basis of their optical spectra and, as a consequence, have not 
been included in the sample of Risaliti et al.\ (1999a). 
The detection of Compton thick matter in objects with broad optical 
emission lines (type I; Guainazzi, Stanghellini, \& Grandi 2003; 
Iwasawa, Maloney, \& Fabian 2002) 
highlights the uncertainties associated with the estimates of the 
occurrence of Compton thick absorption. More examples of AGN that 
do not show any Seyfert signatures in the optical band 
have been reported recently by Maiolino et al.\ (2003): 
{\em Chandra} observations of a small sample of this 
class of optically ``elusive'' nuclei indicate that most of them
are obscured by column densities exceeding 10$^{24}$~cm$^{-2}$.
Their space density is comparable or even higher than that 
of optically selected Seyfert nuclei, implying that the ratio between 
obscured and unobscured AGN is larger than previously estimated.

It is also worth mentioning that high amplitude variability
may play an important role in the source classification. 
Several convincing examples have been discussed by 
Matt, Guainazzi, \& Maiolino (2003) 
where a transition from a reflection dominated spectrum 
to Compton thin and vice versa has been detected on timescales 
of order a few years. The most likely explanation entails strong 
variability of the continuum nuclear source and a Compton thick 
reflector on the parsec scale, possibly associated with the 
absorbing torus. If the nuclear source is switched off, only the 
reflected light is detected. Conversely, if the primary continuum 
source is switched on, the reflected component is not dominant anymore.
Although such an effect should not change, on average, the relative 
ratio between Compton thick and Compton thin absorbers, it adds 
further uncertainties in the estimates of the absorption 
distribution and may also explain the mismatches between optical 
and X-ray classifications if the observations at different 
wavelengths are not simultaneous.

\begin{table}[ht]
%\vskip-24pt
%\vskip -5.2cm
% The first argument to \caption, below, in square brackets
% sends the information to the List of Tables.
% The second argument to \caption, below, in curly brackets,
% contains the text that is printed on the page. 
% This allows the use of indexing commands, for instance, in the table
% caption, without sending that information to the List of Tables.
\caption[Compton thick galaxies observed by {\tt BeppoSAX}]
{\label{comastritab1}
Compton thick galaxies observed by {\em BeppoSAX}}
\begin{tabular*}{\textwidth}{@{\extracolsep{\fill}}lccccc}
\sphline
\em Source Name &\em Opt. Class. &\em $z$ 
&\em $N_{\rm H}$      &\em $L_{\rm 2-10~keV}$  &\em Ref. \cr
                &               &       
& ($10^{24}~{\rm cm}^{-2}$) & ($10^{44}~{\rm ergs~s}^{-1}$) &         \cr 
\sphline
% BeppoSAX results: compilation from Matt et al.\ (2000)
NGC~1068              & Sy2       & 0.0038 & $\ge10$    & $>1$       & (1,2)   \cr %PBLR
Circinus              & Sy2       & 0.0014 & 4.3        & 0.01  & (2,3)   \cr
NGC~6240              & LINER     & 0.0243 & 2.2        & 1.2      & (4)     \cr
Mrk~3                 & Sy2       & 0.0134 & 1.1        & 0.9             & (5,6)   \cr
NGC~7674              & Sy2       & 0.0289 & $\ge10$    &  2               & (7)     \cr %PBLR
NGC~4945              & LINER      & 0.0019 & 2.2$^a$    & 0.03     & (8)     \cr
Tol~0109$-$383        & Sy2$^b$   & 0.0116 & 2.0        & 0.2      & (9)     \cr 
IRAS~09104$+$4109     & QSO2$^c$   & 0.442  & $\ge5^d$   & 80   & (10)       \cr 
NGC~3690              & HII  & 0.011  & $2.5$     & 0.2      & (11)    \cr 
NGC~3281              & Sy2       & 0.0115 & 2.0 & 0.23       & (12)    \cr 
M~51                  & LINER/Sy2 & 0.0015 & 5.6 & 0.0011    & (13)    \cr 
NGC~3079              & LINER/Sy2 & 0.0038 & 10  & 0.01--0.1  & (14)    \cr 
S5~1946$+$708         & RadioGal  & 0.101  & $>$ 2.8  &  $>$ 36  & (15) \cr
PKS~1934$-$63         & RadioGal  & 0.182  & $>$1     &  $>$ 1.9    & (15) \cr
IRAS~20210$+$1121     & Sy2       & 0.0564 & $>10$      & 0.022       & (16) \cr %F2-10=3.0e-13
IC~3639               & Sy2       & 0.0110 & $>10$      & 0.09      & (17) \cr %F2-10=3.47e-13 from Risaliti et al. 99 proc.
%\sphline
% BeppoSAX [from Table 2 of Maiolino et al.\ 1998]
NGC~1386              & Sy2       & 0.0029 & $>1$       & 0.04       & (18)    \cr
NGC~2273              & Sy2       & 0.0062 & $>10$      & 0.08       & (18)    \cr
NGC~3393              & Sy2       & 0.0125 & $>10$      & 0.1        & (18)    \cr
NGC~4939              & Sy2       & 0.0104 & $>10$      & 0.3        & (18)    \cr
NGC~5643              & Sy2       & 0.0039 & $>10$      & 0.045    & (18)    \cr % The only warm scattering in Maiolino; see also R99
MCG$-$05$-$18$-$002   & Sy2       & 0.0056 & $>10$      & 0.05       & (18)    \cr % Reported as IRAS~07145$-$2914 in Risaliti et al. 99
IRAS~11058$-$1131     & Sy2       & 0.0548 & $>10$       & 2.6       & (19)    \cr %F2-10=3.9e-13; see also Ueno et al.00
Mrk~266               & Sy2       & 0.0279 & $>10$       & 0.9       & (19)    \cr %F2-10=5.6e-13
IRAS~14454$-$4343     & Sy2       & 0.0386 & $>10$       & 0.7       & (19)    \cr %F2-10=2.2e-13; see also Ueno et al.00
\end{tabular*}
\begin{tablenotes}
%%%%%%
References --- (1) Matt et al.\ 1997; (2) Guainazzi et al.\ 1999; 
(3) Matt et al.\ 1999; (4) Vignati et al.\ 1999; (5) Cappi et al.\ 1999; 
(6) Matt et al.\ 2000; (7) Malaguti et al.\ 1998; (8) Guainazzi et al.\ 2000; 
(9) Iwasawa et al.\ 2001; (10) Franceschini et al.\ 2000; 
(11) Della Ceca et al.\ 2002; (12) Vignali \& Comastri 2002; 
(13) Fukazawa et al.\ 2001; (14) Iyomoto et al.\ 2001; 
(15) Risaliti, Woltjer, \& Salvati 2003; (16) Ueno et al.\ 1998;
(17) Risaliti et al.\ 1999b; (18) Maiolino et al.\ 1998; 
(19) Risaliti et al.\ 2000. 
\\
%%%%%%
\\
Table notes --- $^{a}$In Schurch, Roberts, \& Warwick (2002), 
$\sim 4$ using {\em Chandra} plus {\em XMM-Newton} observations. 
$^{b}$See also Murayama, Taniguchi, \& Iwasawa (1998) for a 
different optical classification. 
$^{c}$Hyperluminous infrared galaxy with strong narrow emission lines 
(Kleinmann et al.\ 1988). 
$^{d}$Iwasawa et al.\ (2001) find 3.3 using a {\em Chandra} observation. %($>2$). 
\end{tablenotes}
\end{table}
\inxx{captions,table}
\anxx{comptonthick}
\noindent

On the basis of what has been discussed above, it is concluded
that Compton thick absorption is quite common among Seyfert galaxies
in the local universe. Several independent lines of evidence suggest
that mildly and heavily Compton thick sources are likely to dominate
the absorption distribution observed in nearby AGN. The covering
factor of the Compton thick gas must therefore be large.

A compilation of column densities in the Compton thick regime
collected from the literature is reported in
Table~\ref{comastritab1} for objects observed above 10~keV
with {\em BeppoSAX}. The Compton thick nature of sources
observed below 10~keV, inferred from the presence of a strong
iron line, is reported in Table~\ref{comastritab2}.
Due to the lack of coverage above 10~keV, only a conservative
lower limit could be placed on the intrinsic column density.

The majority of the sources are low redshift ($z<0.05$),
relatively low luminosity ($L_X<10^{43}$~ergs~s$^{-1}$)
Seyfert II galaxies. Several interesting exceptions, however,
do exist and will be discussed in the next section.
The $2-10$~keV luminosities are corrected for intrinsic absorption
assuming the best fit spectral parameters reported in the literature.
It is important to point out that for high column densities, the
intrinsic luminosity is strongly dependent upon the precise value of
absorption. For this reason, the luminosities should be considered
only indicative and subject to substantial uncertainties.
This is even more true for those sources for which only a lower
limit for the intrinsic column density is available.
The $2-10$~keV luminosity has been estimated assuming
that 1\% of the intrinsic luminosity is actually observed below
10~keV, due to scattering and or reflection.
This assumption, though reasonable and supported by some observational
evidence (Turner et al.\ 1997), is not always
the rule for Compton thick sources.
Although all efforts have been made to make the list as complete
as possible, some objects could have been missed. Moreover, the
sample is by no means complete and should not be used for statistical
investigations.

\begin{table}[ht]
%\vskip-24pt
%\vskip -5.2cm
% The first argument to \caption, below, in square brackets
% sends the information to the List of Tables.
% The second argument to \caption, below, in curly brackets,
% contains the text that is printed on the page. 
% This allows the use of indexing commands, for instance, in the table
% caption, without sending that information to the List of Tables.
\caption[Compton thick galaxies observed at $E < 10$ keV]
{\label{comastritab2}
Compton thick galaxies observed at $E<10$~keV}
\begin{tabular*}{\textwidth}{@{\extracolsep{\fill}}lccccc}
\sphline
\em Source Name &\em Opt. Class. &\em $z$ 
&\em $N_{\rm H}$      &\em $L_{\rm 2-10~keV}$  &\em Ref. \cr
                &                &       
& ($10^{24}~{\rm cm}^{-2}$) & ($10^{44}~{\rm ergs~s}^{-1}$) &         \cr 
\sphline
% BeppoSAX+ [from Risaliti et al.\ 1999a]
Mrk~1066              & Sy2       & 0.0121 & $>1$       & 0.07      & (20)    \cr %F2-10=2.4e-13 from ASCA/tartarus database
NGC~5005              & Sy2/LINER & 0.0032 & $>1$       & 0.0064   & (20)    \cr %F2-10=3.0e-13
NGC~5347              & Sy2       & 0.0078 & $>1$       & 0.05      & (20)    \cr %F2-10=3.8e-13
IC~5135               & Sy2       & 0.0161 & $>1$       & 0.3       & (20)    \cr %F2-10=5.1e-13; NGC 7130
\sphline
% BeppoSAX+ [from Bassani et al.\ 1999; 5 additional bona-fide Compton thick galaxies base on the location in the 3-D plot - Fig. 1]
% NGC 4968 already included in XMM-Newton sample; 
% NGC 5135 included in Risaliti et al.\ 1999a sample
%%%
NGC~5135              & Sy2       & 0.0137 & $>1$       & 0.075      & (21) \cr %F2-10=2e-13 from Bassani et al. 1999 [Table 2]
NGC~1667              & Sy2       & 0.0152 & $>1$       & 0.01      & (21)    \cr %F2-10=2.6e-14
Mrk~1210              & Sy2       & 0.0135 & $>1$       & 0.8       & (21)    \cr %F2-10=1.3e-12/2.1e-12 [abs/unabs from det. Nh]
Mrk~477               & Sy2/NLSy1  & 0.0378 & $>1$       & 5       & (21)    \cr %F2-10=1.2e-12/1.6e-12 [abs/unabs from det. Nh]
%%%
\sphline
% BeppoSAX+ [from Risaliti et al. 2000-ULIRGs]
% IRAS 20210+1121 already included in Bassani et al.\ 1999 sample; 
% NGC5 5135: already included in Risaliti et al. 1999a
% MOSTLY CANDIDATES COMPTON-THICK GALAXIES; only one indication of Nh in the text
% XMM-Newton: Nh obtained assuming that only the reflection/scattered 
%             component is resent in the XMM-Newton band
% ASCA
ESO~138$-$G1          & Sy2       & 0.0091 & $>1.5$    & 0.33    & (22)    \cr %Nh~2e23 from model E in CB00; L from my calc. 
OQ~$+$208             & RL QSO    & 0.077  & $>$~1   & $\geq$1.4    & (23)    \cr % GPS source; L2-10rf unabs (but reflection dominated)
NGC~6552              & Sy2       & 0.0262 & $>1$       & 1.2     & (24)    \cr %F2-10=8.0e-13; see also Franceschini et al.00
\sphline
% Chandra
IC~2560$^e$           & Sy2       & 0.0097 & $>1^f$     & 0.03       & (25)    \cr % L2-10=2.7e40 % H2O megamaser
NGC~2623            & LINER & 0.0185 & $>1$  &  0.1            & (26) \cr 
NGC~4418            & LINER & 0.0073 & $>1$  &  0.0016         & (26) \cr
UGC~5101            & LINER & 0.0394 & $>1$  &  0.2            & (27) \cr 
\sphline
% XMM-Newton
NGC~4968$^h$          & Sy2       & 0.0099 & $>1$ ?     & 0.15   &  (28)    \cr
IRAS~19254$-$7245     & HII       & 0.062  & $\ge1$    & 10     & (29)    \cr %L,int>e44-45 cgs; see also Risaliti et al. 2000 
IRAS~F12514$+$1027    & QSO/Sy2   & 0.30   & $>1.5$    & $>1.8$     & (30)    \cr % L_intr [but low abs. in XMM band]
\sphline
\end{tabular*}
\begin{tablenotes}
%%%%%%
References --- 
(20) Risaliti et al.\ 1999a; (21) Bassani et al.\ 1999; 
(22) Collinge \& Brandt 2000; (23) Guainazzi et al.\ 2003;  
(24) Risaliti et al.\ 2000; (25) Iwasawa et al.\ 2002; 
(26) Maiolino et al.\ (2003); (27) Ptak et al.\ 2003.
(28) Guainazzi et al.\ 2002; (29) Braito et al.\ 2003; 
(30) Wilman et al.\ 2003.
\\
%%%%%
\\
Table notes ---
$^{e}$~See also Risaliti et al.\ (1999a) for the {\em BeppoSAX} observation. 
$^{f}$~From the iron K$\alpha$ line EW, find $>3$.
$^{h}$~See also Bassani et al.\ (1999; candidate Compton thick galaxy).  
\\
\end{tablenotes}
\end{table}
\inxx{captions,table}
\anxx{comptonthick}
\noindent

\section{Compton Thick AGN and the XRB}
\label{comastrisecunified}

The amount of obscuring gas in Seyfert galaxies and quasars, 
and the distribution of column density as a function of redshift
and luminosity, are key ingredients in the X-ray background synthesis 
models. First predicted by Setti \& Woltjer (1989) and elaborated 
on with an increasing degree of detail by several authors 
(Madau, Ghisellini, \& Fabian 1994; Comastri et al.\ 1995; 
Gilli, Salvati, \& Hasinger 2001), the XRB spectral energy density
is due to the integrated contribution of highly obscured AGN.

Although a detailed description of population synthesis models for 
the XRB is beyond the purposes of the present chapter, it might be 
useful to summarize briefly the main features (see Comastri 2001 
for more details). The basic recipe to ``cook'' the so-called baseline 
model is a rather straightforward three-step approach.
(1) Assume as a template for unobscured (type I) sources the average 
spectrum of nearby Seyfert I galaxies above 2~keV, parameterized as 
a steep ($\Gamma=1.9$) power law plus a reflection component from 
a face-on disk and a high energy cutoff at about 300~keV.
(2) Add to the template a distribution of column densities with 
$\log N_H$ in the range $21-25$~cm$^{-2}$ to model the obscured 
(type II) AGN population. (3) Fold both the type I and type II 
spectra with an evolving X-ray luminosity function (XLF) with 
best fit parameters determined from soft X-ray surveys and 
appropriate for unobscured AGN.

The column density distribution is varied in both shape and normalization 
until a good fit to the XRB spectrum, source counts in different energy 
bands, and redshift distributions at different limiting fluxes 
is obtained. While at first glance it seems that the only parameters 
that are free to vary are those related to the absorption distribution, 
it must be stressed that the best fit values adopted for the
evolving XLF and spectral templates are also subject to non-negligible 
uncertainties that are not taken into account
in the baseline model. In particular, the assumptions concerning 
the XLF evolution of the obscured population, though in line with 
a strict version of the AGN unified scheme, do not have any 
observational support. It is thus remarkable that a such a model 
was able to reproduce all of the observational constraints available 
in the pre-{\em Chandra}/{\em XMM-Newton} era.
It is also worth remarking that the baseline model (as well as all of
the models proposed so far) is purely X-ray based: the optical 
appearance of an obscured source may not necessarily be that 
of an optically selected type II AGN.

%
% FIGURE 3
%
\begin{figure}[tbh]
\vskip -1.3cm
\centerline{\includegraphics[width=0.8\textwidth]{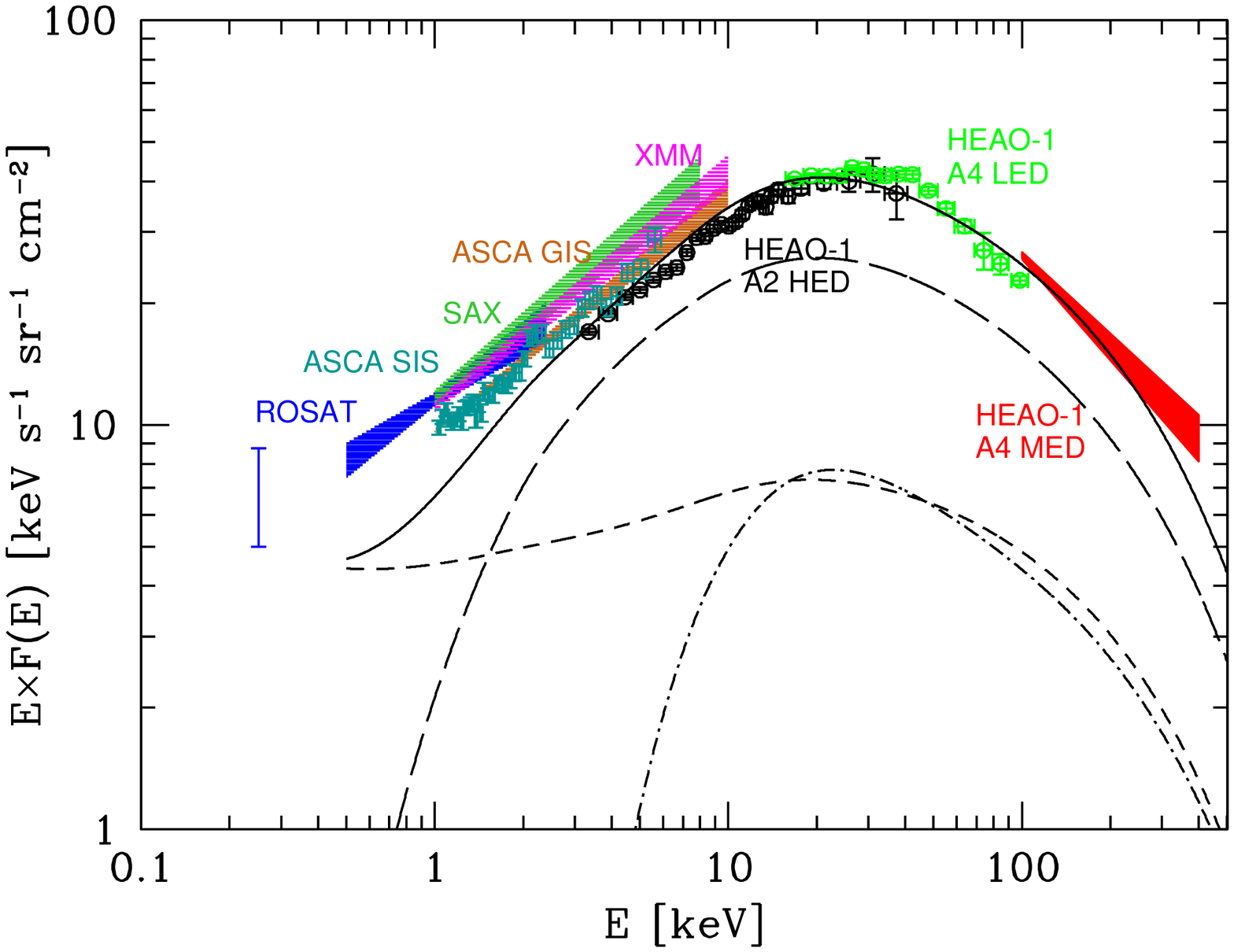}}
\vglue-2.15cm
\centerline{\includegraphics[width=0.8\textwidth]{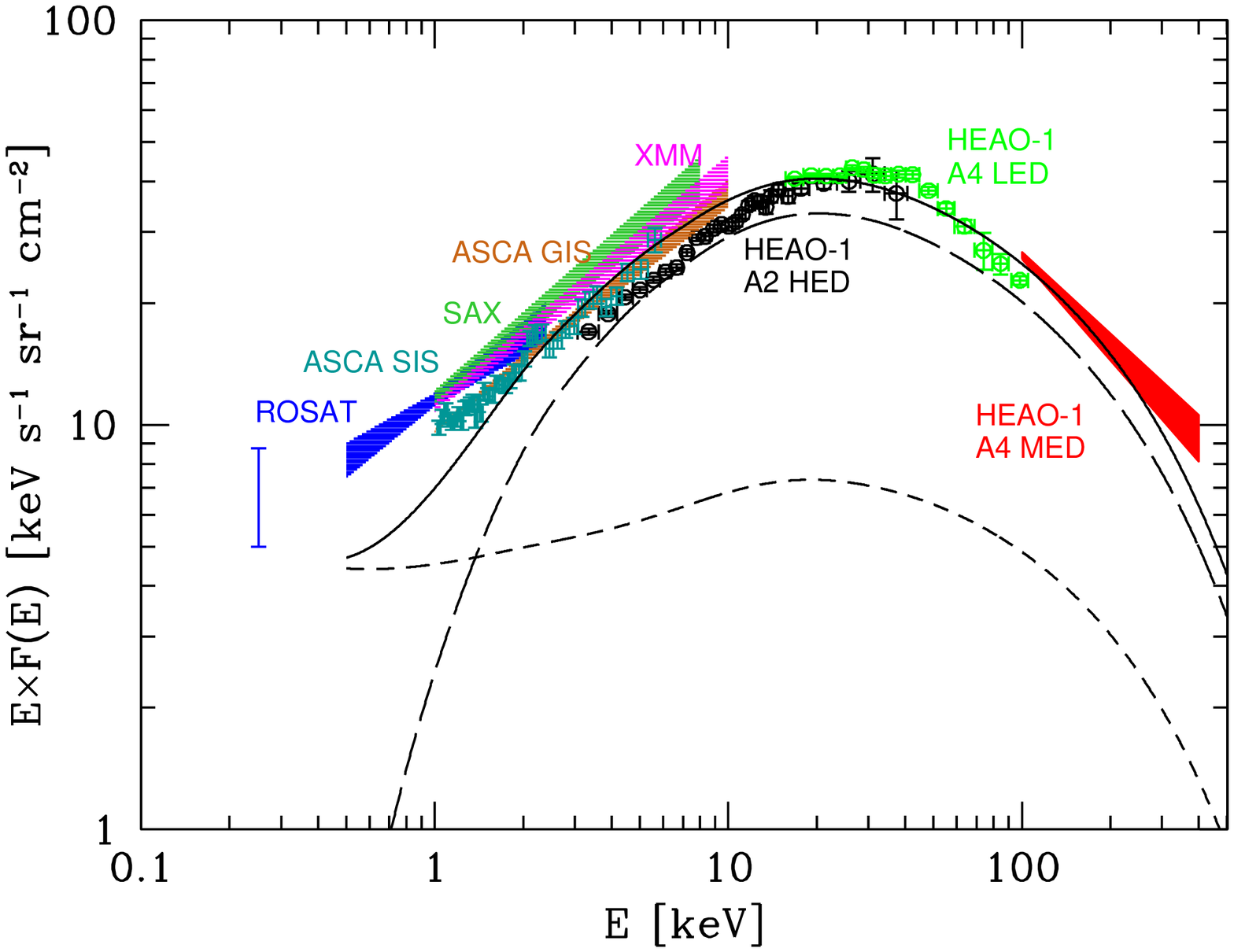}}
\vskip -1.0cm
\caption{{\em (Top panel)} AGN contribution {\em (solid line)} to the
XRB spectral energy density, as measured by different instruments
{\em (labeled)}. Contributions of unobscured {\em (dashed line)},
Compton thin {\em (long-dashed line)}, and Compton thick
{\em (dot-dashed line)} AGN are also shown. Labels refer to XRB
measurements obtained by different satellites.
{\em (Bottom panel)} Same as above, but without Compton thick AGN.}
\label{comastrifig3}
\end{figure}

The integrated contributions of unobscured and obscured AGN, the
latter split into Compton thin and Compton thick AGN, are
reported in Figure~\ref{comastrifig3} (adapted from Comastri et al. 2001).
Even though such a model is not able to reproduce some of the
recent observational constraints emerging from deep {\em Chandra}
and {\em XMM-Newton} surveys---in particular, the observed
redshift distribution---it can be considered representative
of most of the population synthesis models, at least regarding
the contribution of Compton thick sources to the broadband XRB
spectrum (see, for example, Fig.~5 in Wilman \& Fabian 1999).
Indeed, in the synthesis model of Ueda et al.\ (2003), where a
more satisfactory description of the recent observational constraints
is obtained, the integrated contribution of mildly Compton thick
sources is very similar to that shown in Figure~\ref{comastrifig3}.

Not surprisingly, mildly Compton thick sources provide a 
non-negligible contribution only around the peak energy of the 
XRB spectrum, while it is clear that most of the energy density 
is accounted for by obscured, Compton thin AGN. 

Given that the overall shapes of Compton thin and Compton thick AGN, 
once convolved with the evolving XLF, are very similar, 
it seems reasonable to argue that the XRB spectrum could be 
equally well fitted without invoking Compton thick absorption.
Indeed, this appears to be the case: an acceptable fit to the 
XRB spectrum can be obtained just by rescaling the absorption 
distribution of Compton thin AGN (see Fig.~\ref{comastrifig3}).

%
% FIGURE 4
%
\begin{figure}[ht]
\centerline{\includegraphics[width=0.9\textwidth]{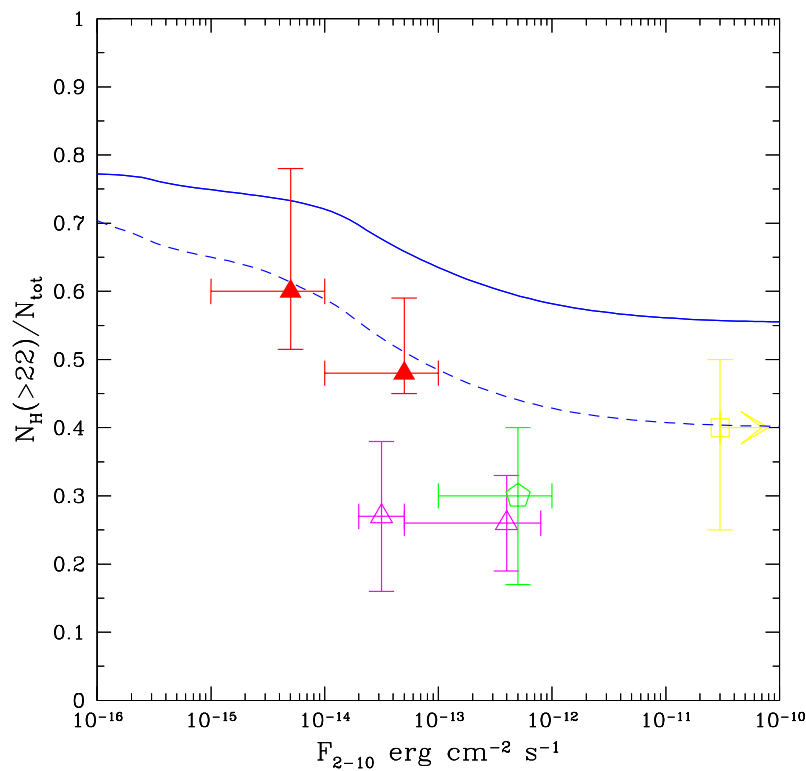}}
\vskip -0.25cm
\caption{Expected fraction of sources with absorption column 
densities larger than 10$^{22}$~cm$^{-2}$ vs. $2-10$~keV flux
for a model without {\em (solid line)} and with {\em (dashed line)} 
Compton thick sources. Points and associated error bars correspond
to the observed fraction of absorbed sources in different X-ray 
surveys: open square at bright X-ray fluxes is from the 
{\em HEAO1-A2} AGN sample of Piccinotti et al.\ (1982), open 
hexagon is from the {\em ASCA} Large Sky Survey (Akiyama et al.\ 2000), 
open triangles are from {\em XMM-Newton} observations (Piconcelli et al.\ 2002),
and filled triangles are from {\em Chandra} observations (Brusa 2004).}
\label{comastrifig4}
\end{figure}

However, the relative fraction of absorbed ($N_H>10^{22}$~cm$^{-2}$) 
sources as a function of the $2-10$~keV flux is significantly 
overestimated over the entire range of fluxes (solid line in 
Fig.~\ref{comastrifig4}). While the model that includes 
Compton thick absorption does fit the observations at both very 
bright and faint fluxes better (dashed line in Fig.~\ref{comastrifig4}), 
the absorption distribution measured by shallow, large-area 
{\em ASCA} and {\em XMM-Newton} surveys is not well matched.
Such a discrepancy, together with the redshift distribution of obscured
AGN, is challenging the standard model for the XRB, at least 
in its simplest form, calling for some revision of the basic 
assumptions. A detailed comparison of XRB modeling versus 
observations is beyond the purposes of this chapter.
The bottom line of the exercise described above is that 
Compton thick absorption cannot be neglected in building up a 
synthesis model for the XRB. 

On the other hand, it is also possible to fit the XRB spectrum 
with a model where the contribution of heavily Compton thick AGN 
is dominant. This possibility has been put forward by 
Fabian et al.\ (1990), who showed that the $10-300$~keV background 
intensity can be accounted for by the integrated contribution of 
luminous sources at $\langle z \rangle \sim 1.5$ with the 
characteristic Compton ``reflection'' spectrum 
(see Fig.~\ref{comastrifig2}). The most interesting feature of 
this model is the natural explanation of the XRB 30~keV peak in 
terms of a physical process that is basically driven by the Thomson 
cross-section, $\sigma_T$. The integrated spectrum of Compton 
reflection dominated sources is, however, too hard and does not 
fit the data below 10~keV; moreover, the source counts 
in the $0.3-3.5$~keV band (the only available at that time) 
were strongly underestimated (Comastri 1991; Terasawa 1991).

\begin{figure}[ht]
\centerline{\includegraphics[width=3in, angle=-90]{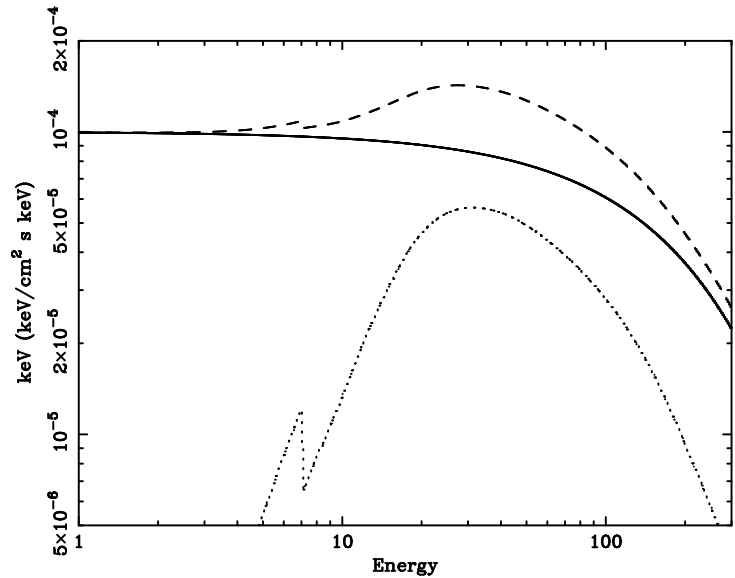}}
\caption{Long dashed curve is the X-ray spectrum after reflection from 
cold thick matter subtending $2\pi$~sr at the illuminating continuum 
source---a power law with $\Gamma=2$ plus an exponential 
cut-off at 300~keV {\em (solid curve)}. Dotted curve shows the 
reflected component.
}
\label{comastrifig5}
\end{figure}

Although such an extreme hypothesis is ruled out by the data, 
it has been pointed out that the 30~keV peak can be better 
reproduced including a  ``reflection'' component (Gilli et al.\ 2001; 
Ueda et al.\ 2003) with a covering factor of about $2\pi$ 
(Fig.~\ref{comastrifig5}) in the spectral templates of both 
unobscured and obscured AGN. 

%
% FIGURE 5
%

On the basis of these considerations, it is safe 
to conclude that absorption and reflection from Compton thick 
matter need to be included in the high energy spectra of 
the sources responsible for the hard XRB. While there is compelling 
evidence (see \S~\ref{comastriseclocal}) for a numerous population 
of Compton thick AGN in the nearby universe, their space density 
at cosmological distances is still basically unknown.

\section{Distant Obscured AGN}

Thanks to deep and medium-deep {\em Chandra} and {\em XMM-Newton} 
surveys (Giacconi et al.\ 2002; Alexander et al.\ 2003; 
Hasinger et al.\ 2001), it is now possible to search for obscured 
AGN at much larger distances. The determination of the intrinsic 
column density requires that the source redshift be known. 
Extensive campaigns of optical identifications have started to 
produce sizable samples of spectroscopically identified sources 
(Barger et al.\ 2002, 2003; Mainieri et al.\ 2002; 
Szokoly et al.\ 2004). 

Most of the hard X-ray selected sources are obscured by column 
densities $10^{21}<N_H<10^{23}$~cm$^{-2}$, with unabsorbed hard X-ray 
luminosities in the range $10^{42}-10^{44}$~ergs~s$^{-1}$. These values 
are typical of Compton thin Seyfert galaxies in the nearby universe.
Obscured AGN with quasar-like luminosities and column densities 
extending to the Compton thick regime are foreseen by the synthesis 
models of the XRB based on the AGN unified scheme 
(\S~\ref{comastrisecunified}).
Although it has been predicted (Fabian, Wilman, \& Crawford 2002) 
that deep (of order a million seconds) {\em Chandra} exposures 
should detect several tens of distant, optically faint, Compton thick
quasars, the observational evidence for such sources is rather 
scanty. Indeed, only a handful of obscured sources with quasar 
luminosities ($L_X>10^{44}$~ergs~s$^{-1}$) have been discovered so 
far (e.g., Stern et al.\ 2002), with only a marginal indication of 
column densities in excess of 10$^{24}$~cm$^{-2}$ in a few objects
(Norman et al.\ 2002; Wilman et al.\ 2003).
The low signal-to-noise X-ray spectrum of the 
high redshift ($z=3.7$) quasar {\em Chandra} Deep Field-South-202 
(CDF-S-202) can be equally well fitted by a transmission model 
with $N_H\simeq 8\times 10^{23}$~cm$^{-2}$ and by a reflection 
dominated continuum implying a heavily Compton thick source 
(Norman et al.\ 2002). In both cases, the iron line EW is of order 
1~keV, though the detection is significant only at the 2$\sigma$ 
level. The optical spectrum is dominated by narrow lines with an 
average FWHM of order 1500~km~s$^{-1}$ and no continuum, 
as expected for a type II quasar.

Detections of Compton thick sources with quasar luminosities
also have been reported from X-ray observations of
hyperluminous infrared galaxies (HyLIRGs). The power source 
responsible for an observed rest-frame $1-1000~\mu$m luminosity
in excess of $10^{13}$~L$_{\odot}$ in these objects is thought 
to be a mixture of obscured star formation and AGN activity.

The best evidence for a deeply buried 
($N_H\geq 5\times 10^{24}$~cm$^{-2}$), luminous quasar
($L_{10-100~{\rm keV}}\simeq 1.7\times 10^{46}$~ergs~s$^{-1}$)   
comes from broadband 
{\em BeppoSAX} observations to $\sim100$~keV
of the HyLIRG IRAS~09104$+$4109, the most luminous 
object in the universe at $z<0.5$ (Franceschini et al.\ 2000). 
Unfortunately, the {\em BeppoSAX} sensitivity above 10~keV 
does not allow other deeply buried HyLIRGs to be detected in
this way. However, more sensitive {\em XMM-Newton} observations 
below $8-10$~keV provide some indirect evidence of Compton thick 
obscuration in two additional X-ray luminous HyLIRGs 
(Wilman et al.\ 2003). 

Despite the above results that indicate that Compton thick quasars
do indeed exist, the present observations suggest that their 
relative fraction among the obscured AGN population at high 
redshifts is likely to be much lower than in the nearby universe. 

The disagreement between optical and X--ray classifications 
(already discussed in \S~\ref{comastriseclocal}) 
may be responsible for the difficulties encountered in the 
search for, and classification of, distant obscured sources.
Indeed, one of the most surprising findings of the {\em Chandra} 
and {\em XMM-Newton} surveys is the discovery of a 
sizable number of relatively bright X-ray sources 
spectroscopically identified with early-type ``normal'' galaxies 
without any obvious signatures of nuclear activity in their optical 
spectra (Mushotzky et al.\ 2000; Fiore et al.\ 2000; 
Barger et al.\ 2001a,b; Hornschemeier et al.\ 2001; 
Giacconi et al.\ 2001)
It should be pointed out that the presence of relatively bright 
X-ray sources in the nuclei of optically dull galaxies was first 
recognized by {\em Einstein} observations (Elvis et al.\ 1981).
The very nature of X-ray Bright Optically Normal Galaxies 
(hereafter, XBONG) is still a matter of debate. Multiwavelength 
observations of a small sample of 10 ``bona fide'' sources belonging 
to this class were discussed by Comastri et al.\ (2002a). 
The high X-ray luminosities and the relative ratios between 
X-ray, optical, and radio fluxes strongly suggest that 
nuclear activity is present in almost all the XBONG.
A detailed investigation of what can be considered
the prototype of this class of object (P3; Comastri et al.\ 2002b) 
indicates that a heavily obscured (possibly Compton thick) AGN 
is the most likely explanation. Further evidences favouring the 
Compton thick absorption have been presented by Comastri et al.\ (2003).

However, such a possibility is not unique,
and alternative solutions could still be viable.
For example, the broadband properties of two XBONG
discussed by Brusa et al.\ (2003) are more consistent 
with those of a BL Lacertae object and a small group of 
galaxies, respectively. It also has been pointed out 
(Moran, Filippenko, \& Chornock 2002; see also Moran, this volume) 
that optical spectroscopy sometimes can be inefficient in 
revealing the presence of an AGN, 
especially at faint fluxes, when only low quality optical spectra 
are in hand. The nuclear emission lines easily could be diluted 
by the host galaxy starlight. Indeed, AGN features in the optical 
band were found in three bright candidate XBONG, serendipitously 
discovered by {\em XMM-Newton} (Severgnini et al.\ 2003),
thanks to high signal-to-noise ratio spectroscopic observations 
that were able to separate the nuclear spectrum from that of the 
host galaxy. The X-ray and optical properties 
could then be explained by moderately luminous 
($L_X\sim 10^{42-43}$~ergs~s$^{-1}$), mildly obscured 
($N_H\sim 10^{22}$ cm$^{-2}$) AGN hosted by luminous galaxies.
While it is almost certain that XBONG are powered by
accretion onto a supermassive black hole, it remains to be proved 
that Compton thick absorption is ubiquitous among these sources.

%
% FIGURE 6
%
\begin{figure}[ht]
\centerline{\includegraphics[width=\textwidth]{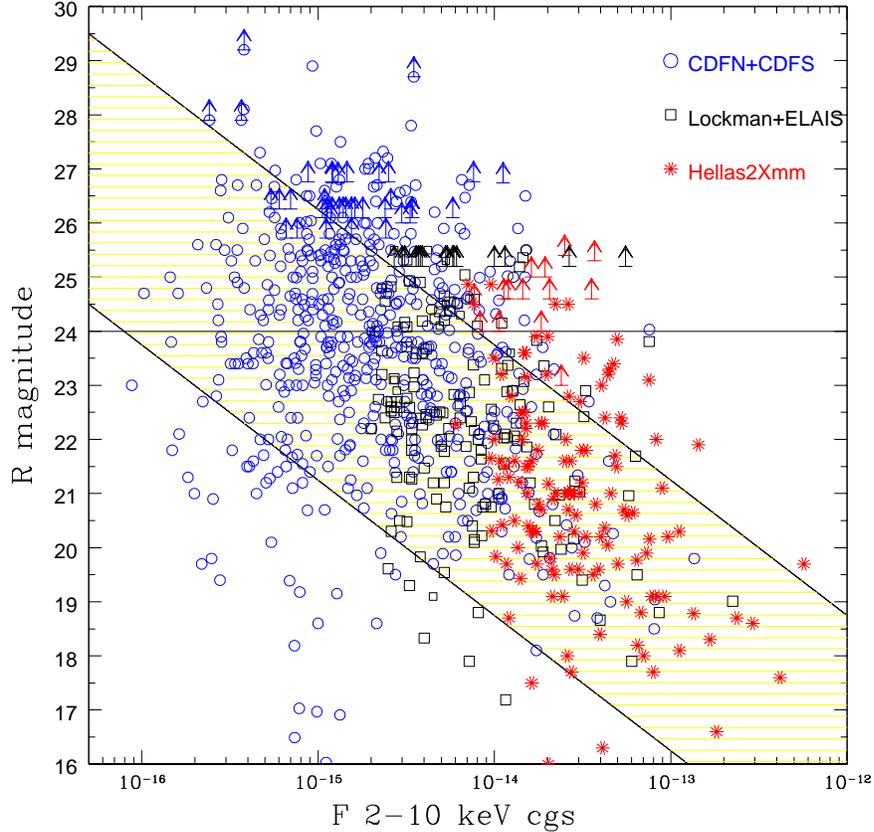}}
\caption{Optical $R$ magnitude vs. $2-10$~keV flux
for hard X-ray selected sources in deep {\em Chandra} and
medium deep {\em XMM-Newton} surveys {\em (labeled)}.}
\label{comastrifig6}
\end{figure}

High redshift Compton thick AGN could be hiding among the relatively
large number of X-ray sources that have remained so far unidentified 
due to their extremely faint optical counterparts. The ratio between 
X-ray flux and optical magnitude (defined as $\log F_X/F_{opt} = 
\log F_X + 5.5 + R/2.5$, see Maccacaro et al.\ 1988; 
Barger et al.\ 2002; McHardy et al.\ 2003) is considered to be 
a fairly reliable indicator of the X-ray source classification.
Indeed, the distribution of $F_X/F_{opt}$ values of 
most of the spectroscopically identified X-ray selected AGN 
from {\em ROSAT} (Hasinger et al.\ 1998), {\em ASCA} 
(Akiyama et al.\ 2003), {\em Chandra} (Giacconi et al.\ 2001) 
and {\em XMM-Newton} (Mainieri et al.\ 2002; Fiore et al.\ 2003) 
surveys fall within $-1<\log F_X/F_{opt}<1$.

\begin{figure}[ht]
\centerline{\includegraphics[width=4in, angle=0]{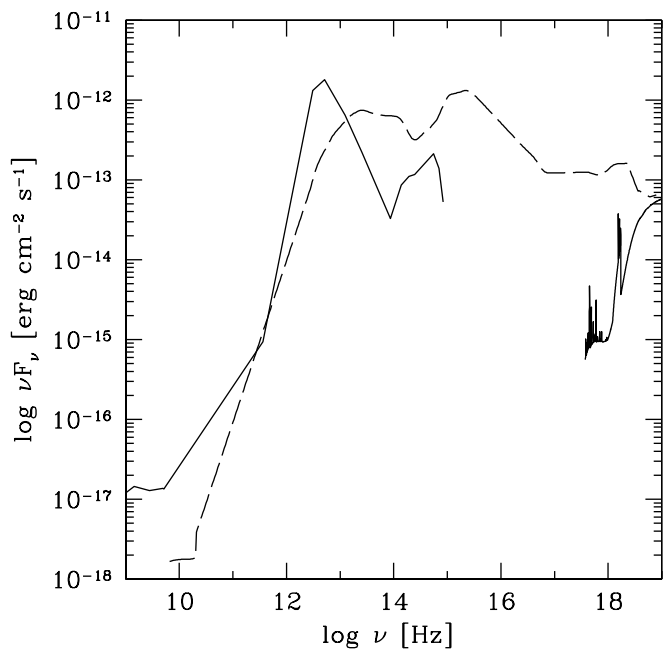}}
\caption{Spectral energy distribution of NGC~6240 {\em (continuous line)}
and of the average spectrum of radio-quiet quasars (Elvis et al.\ 1994;
{\em dashed line}) showing a significantly different behavior at both
UV/optical and X-ray frequencies.}
\label{comastrifig7}
\end{figure}

A sizable fraction (of order 20\%) of  hard X-ray selected sources 
are characterized by $\log F_X/F_{opt}>1$ (Fig.~\ref{comastrifig6}). 
Obscured accretion seems to provide the most likely explanation 
for high $F_X/F_{opt}$ values. Such a possibility 
was confirmed by deep VLT spectroscopy of a few X-ray bright sources 
that turned out to be high redshift, type II AGN (Fiore et al.\ 2003). 
It has been further suggested (Comastri et al.\ 2003)
that high, and even extreme, values of $\log F_X/F_{opt}$ 
could be easily reproduced by shifting the broadband 
spectral energy distribution (SED), including the host galaxy 
starlight, of prototype Compton thick 
sources (i.e., NGC~6240 and IRAS~09104$+$4109) 
over a range of redshifts ($z=0.5-1.5$). 
This range should be considered indicative, as the
optically faintest sources, especially those not detected 
in the $R$ band images, could be at greater distances. 
The combined effect of the
$K$-corrections on the observed optical and hard X--ray fluxes due to 
the very shape of the NGC~6240 SED (Fig.~\ref{comastrifig7}) is 
responsible for the non--linear behavior of the $F_X/F_{opt}$ ratio
as a function of redshift.

%{\em ZZZ Does this include a host galaxy? That would probably have
%to be at higher redshifts to be missed. ZZZ}
%
% FIGURE 7
%

%
% FIGURE 8
%
\begin{figure}[ht]
\centerline{\includegraphics[width=3in, angle=-90]{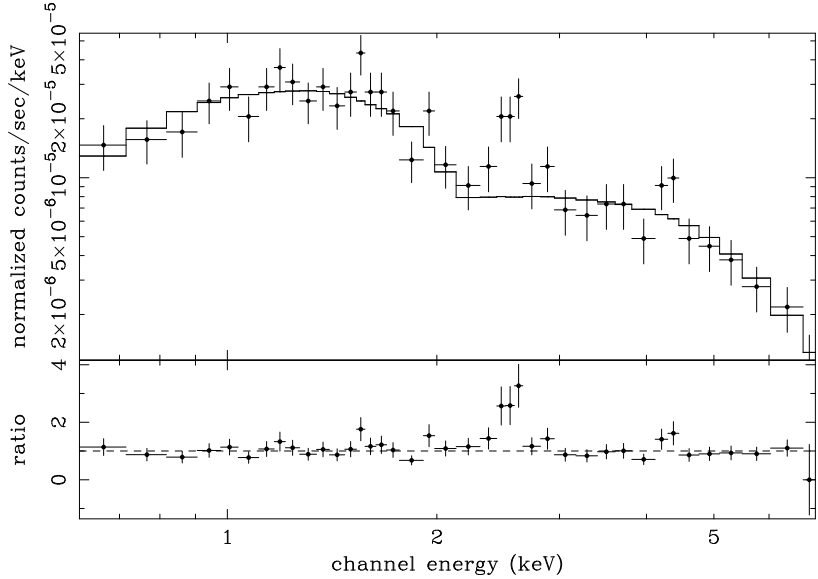}}
\caption{Simulated spectrum of a reflection dominated 
$F_{2-10~{\rm keV}}\sim 10^{-15}$~ergs~cm$^{-2}$~s$^{-1}$ source 
at $z=1.5$, as would be seen with an ultradeep 10~Ms {\em Chandra} 
observation. Input parameters are a flat power law with $\Gamma$=1 
plus an iron line with an observed EW of 600~eV. The line feature 
is clearly seen in the residuals (bottom panel) of a single power 
law fit.}
\label{comastrifig8}
\end{figure}

In order to establish whether the high values of X-ray to optical 
flux ratios are due to Compton thick obscuration, good quality 
X-ray spectra are needed to search for the characteristic 
absorption and reflection features of thick matter.
In particular, the detection of a strong redshifted iron 
K$\alpha$ line would provide strong evidence for Compton thick gas 
and, most importantly, would allow a reliable redshift measurement 
to be obtained for objects with optical magnitudes 
beyond the spectroscopic limit of large optical telescopes.
Unfortunately, the X-ray counting statistics, even in the deep 
{\em Chandra} fields, are not such that they can properly address 
this issue, possibly explaining the small fraction ($\sim 7$\%) 
of CDF-N sources showing iron K$\alpha$ emission lines 
(Bauer et al.\ 2003). 

A systematic search for strong (EW~$\sim 1-2$~keV), redshifted   
iron lines among relatively faint 
($F_{2-10~{\rm keV}}\sim 10^{-15}$~ergs~cm$^{-2}$~s$^{-1}$) 
hard X-ray sources would require extremely deep (of order 10~Ms) 
{\em Chandra} observations (Fig.~\ref{comastrifig8}). These are 
not limited, at least in the inner part of the detector, 
by source confusion, thanks to the good point spread function (PSF).

\begin{figure}[ht]
\centerline{\includegraphics[width=4in]{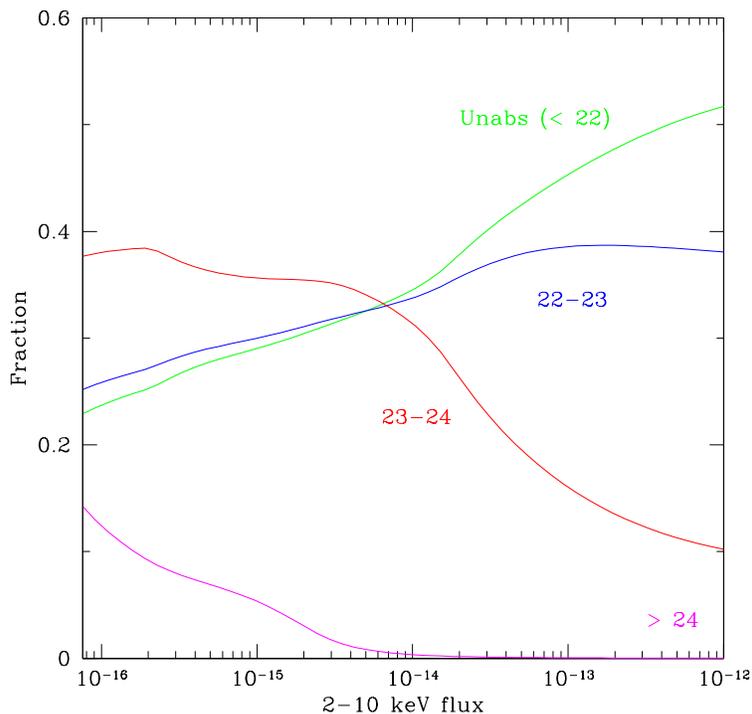}}
\caption{Relative fraction of sources with different column 
densities (labeled) vs. X-ray flux.}
\label{comastrifig9}
\end{figure}

The fraction of X-ray sources versus intrinsic column density 
predicted by current synthesis models is shown in 
Figure~\ref{comastrifig9}. The relative number of sources with 
column densities larger than 10$^{24}$~cm$^{-2}$ starts to increase 
below fluxes of order 10$^{-15}$~ergs~cm$^{-2}$~s$^{-1}$, reaching 
a fraction of order 10\% of the entire X-ray source population
around the limit of the deepest {\em Chandra} survey 
($F_{2-10~{\rm keV}}\sim 2\times 10^{-16}$~ergs~cm$^{-2}$~s$^{-1}$). 
A  similar fraction is also predicted by the Ueda et al.\ (2003) 
model, which is based on quite different prescriptions for the 
evolution of the luminosity function.   

%
% FIGURE 9
%

The small area covered with good spatial resolution by a single
ACIS field would imply that at least a few pointings are needed 
to collect a sizable number of sources for statistical investigations. 
Such a project, although technically feasible, appears to be 
extremely challenging in terms of observing time. The search for 
Compton thick AGN through an X-ray selection based on the detection 
of a strong iron line appears to be rather inefficient
and probably not worth pursuing.

\section{Black Hole Mass Density of Compton Thick Sources}

The local density in black holes grown by accretion with efficiency
$\eta$ can be computed using the integral argument originally proposed 
by Soltan (1982). An excellent review of the method, along with 
a detailed discussion of the most recent determinations 
of the black hole mass density, is presented by Fabian (2004).
Here I limit myself to estimating the black hole mass accreted
during the Compton thick phase, $\rho_{\bullet}(CT)$,
under the hypothesis that the Compton thick source contribution 
to the XRB is that reported in Figure~\ref{comastrifig3}. 

The unabsorbed $2-10$~keV X-ray flux that enters into the 
computation of $\rho_{\bullet}(CT)$ is obtained by assuming 
$\Gamma$=2 and $N_H=3\times 10^{24}$~cm$^{-2}$,
\begin{equation}
\rho_{\bullet}(CT)={k_{bol}\over \eta c^2}~(1+\langle z \rangle)~8.2\times 
10^{-18}~{\rm ergs~cm}^{-3} \,.
\end{equation}
The calculation of $\rho_{\bullet}(CT)$ then requires the knowledge 
of the bolometric correction term, the average redshift of the sources,
and the accretion efficiency. For the latter, a standard value of
$\eta=0.1$ is assumed, while for the average redshift, the latest 
results of optical identifications suggest $\langle z \rangle \sim 1$
(e.g., Gilli 2004). 
The bolometric correction from the unabsorbed $2-10$~keV luminosity 
to the total luminosity is known to be a function of the X-ray 
luminosity. Correction factors in the range $k_{bol} \sim 10-20$  
appear to be appropriate for low luminosity Seyfert-like objects 
(Fabian 2004), to be compared with values in the range 
$30-50$ for high luminosity unobscured quasars (Elvis et al.\ 1994). 
As far as the two well studied prototype Compton thick sources
NGC~6240 and IRAS~09104+4109 are concerned, a direct calculation 
from broadband literature data, including the contribution
of the host galaxy, provides remarkably similar results: 
$k_{bol} \sim 10$. A more precise estimate of the bolometric 
correction factor as a function of the X--ray properties is not 
possible with the present data. {\em Spitzer Space Telescope} 
observations of well defined samples of obscured AGN will 
provide a significant step forward in this direction. 

Taking the above estimates for $k_{bol}$, $\eta$, and $\langle z \rangle$  
at face value, the black hole mass density 
accreted in the Compton thick phase is about 
$3\times 10^4$~M$_{\odot}$~Mpc$^{-3}$. 
This value could be slightly higher (up to 
$5\times 10^4$~M$_{\odot}$~Mpc$^{-3}$) if the intrinsic spectral 
parameters are different. Based on the same argument, but following 
a slightly different approach, Fabian (2004) concluded that 
$\rho_{\bullet}(CT)<10^5$~M$_{\odot}$~Mpc$^{-3}$. 
Unless the estimates of $k_{bol}$, the average redshift, and the 
relative contribution of Compton thick sources to the XRB are revised 
upwards (which seems rather unlikely), 
they only account for some 10\% of the black hole mass density 
grown by accretion. Thus, most of the black hole mass density is 
due to less obscured Compton thin sources and unobscured
quasars (see Cowie \& Barger, this volume).

\section{Contribution to the Far-IR Background}

The X-ray and UV energy density absorbed by the torus of type II
AGN is reemitted at longer wavelengths, most likely in the far-IR.
It is thus plausible that obscured AGN may significantly 
contribute to the extragalactic background light in this band. 
In principle, such a calculation could be performed following 
the guidelines discussed in \S~\ref{comastrisecunified} 
for the synthesis of the XRB spectrum. In practice, such an 
approach is unfeasible, mainly because the average AGN spectral 
shape at long wavelengths is still rather uncertain, both from 
an observational and from a theoretical point of view.

In order to overcome these difficulties, a simple integral argument
can been employed to compute the AGN contribution to the 
far-IR background. The reprocessed luminosity is 
estimated from the hard X-ray luminosity using the XRB intensity 
as an upper limit of the total energy output (Fabian \& Iwasawa 1999).
If one further assumes reasonable values for the effective 
temperature of the reprocessing dust and the cosmological evolution of
the luminosity function, as determined from X-ray surveys 
(Almaini, Lawrence, \& Boyle 1999; Brusa, Comastri, \& Vignali 2001; 
Risaliti, Elvis, \& Gilli 2002), it is also possible to estimate 
the spectral shape of the long wavelength background due to X-ray 
emitting AGN. 

The different approaches come to a similar result: obscured AGN 
contribute from a few percent up to a maximum of $10-15$\% at 
wavelengths longer than about $100\mu$m, being possibly higher 
in the mid-IR ($15-60\mu$m; Risaliti et al.\ 2002). 
These findings are in relatively good agreement with observations 
in the infrared and submillimeter bands. The cross-correlation 
of {\em ISO} and {\em XMM-Newton} sources in the Lockman Hole 
(Fadda et al.\ 2002) indicates an AGN contribution 
of about 15-20\% at $15\mu$m (see also Alexander et al.\ 2002), 
which is similar to that estimated by Matute et al.\ (2002) from 
the optical identifications of the {\em ELAIS} field.
In the submillimeter, Barger et al.\ (2001c) found that the
ensemble of X-ray sources in the 1~Ms exposure of the CDF-N
with SCUBA observations contribute about 15\% of the
extragalactic light at 850$\mu$m.
According to the deepest investigation so far carried out using 
the 2~Ms CDF-N data (Alexander et al.\ 2003), many SCUBA sources 
are rather faint at high energies, suggesting that most of the 
submillimeter background is due to stellar processes.

\begin{figure}[ht]
\vskip -0.5cm
\centerline{\includegraphics[width=5in]{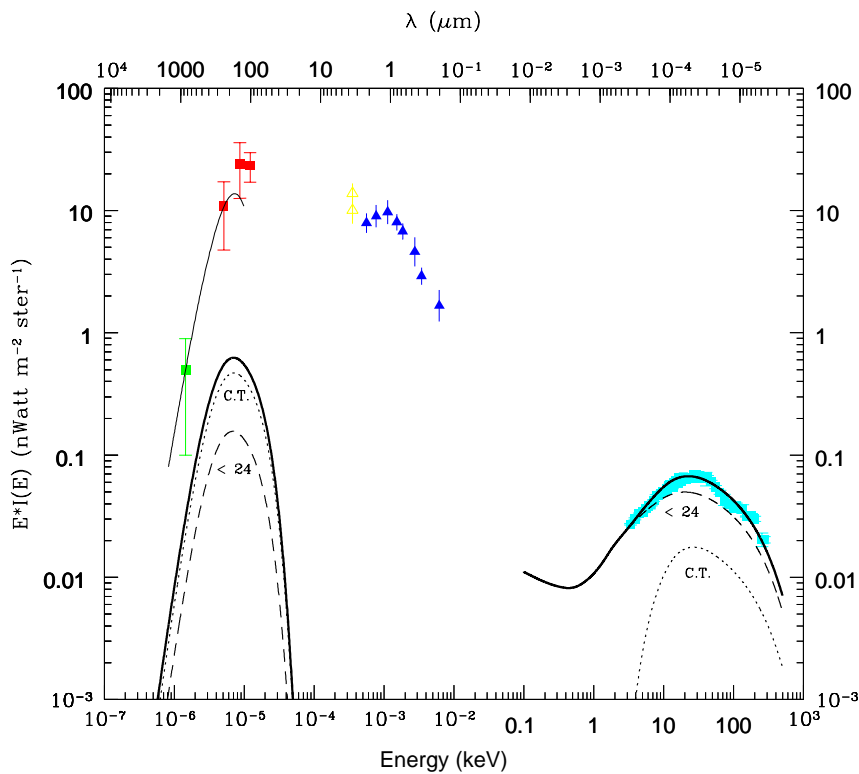}}
\vskip -1.5cm
\caption{Model predicted AGN contribution at far-IR and X-ray
wavelengths (solid thick line). Dashed line represents
contribution of Compton thin sources, while dotted line that
of Compton thick sources. Data points are from a compilation of
measurements of the extragalactic background intensity from
Fixsen et al.\ (1998; submillimeter data and best fit curve---thin
black curve), Lagache et al.\ (2000; far-IR data), Dwek \& Arendt
(1998; DIRBE mid-IR data), Pozzetti et al.\ (1998; optical
and near-IR data) and Marshall et al.\ (1980; X-ray data).}
\label{comastrifig10}
\end{figure}

An attempt to estimate the contribution of Compton thick sources 
to the long wavelength background is discussed by Brusa et al.\ (2001).
Assuming the XRB model of Comastri et al.\ (1995) and the observed 
correlations between the hard X-ray and far-IR luminosities 
of bright nearby AGN as a function of the absorbing column density, 
these authors concluded that, although not energetically dominant, 
the most important contribution to the far-IR background 
comes from Compton thick sources (Fig.~\ref{comastrifig10}).

If anything, heavily obscured Compton thick sources appear to 
be the most favored class of AGN to be detected in the far-IR/submillimeter
bands. The issue of the nature of the far-IR background will greatly 
benefit from upcoming {\em Spitzer Space Telescope} observations.  

%
% FIGURE 10
%

\section{Conclusions and future prospects}
 
Different sets of X-ray observations and theoretical arguments 
indicate that the reprocessing of high energy radiation 
by Compton thick matter in the circumnuclear regions of AGN 
is very common, if not ubiquitous. 
Unfortunately, even the most sensitive hard X-ray surveys 
carried out with the {\em Chandra} and {\em XMM-Newton} 
observatories turn out to be rather inefficient for 
searching for the typical signatures of thick matter. 
Further observations, though useful in several respects,
will not provide the breakthrough in this direction. 
In fact, according to what can be considered a solid model 
for the XRB, only a small fraction of the Compton thick 
population could be detected by present X-ray surveys below 
$8-10$~keV.

A mission capable of exploring the hard X-ray sky in the 
$10-70$~keV band with focusing/imaging instruments able to 
reach fluxes of order $10^{-14}$~ergs~cm$^{-2}$~s$^{-1}$ 
would provide a quantum leap forward. As an order of magnitude 
estimate, the three dex jump in the limiting flux 
would be similar to that achieved by {\em BeppoSAX} and 
{\em ASCA} in the $2-10$~keV band with respect to the 
{\em HEAO1-A2} survey of Piccinotti et al.\ (1982).
In other words, such a gain in sensitivity would push the resolved 
fraction of the hard ($10-70$~keV) XRB from a negligible 1\% 
to about $30-40$\%. Sensitive, high energy surveys 
will open up a large volume of discovery space 
that so far only has been touched by {\em BeppoSAX} observations. 
It is important to stress that the scientific output of these 
observations would be even more rewarding than a mere improvement 
in the limiting flux, as the $10-70$~keV band fully encompasses 
the peak of the XRB energy density at 30~keV.

The X-ray Evolving Universe Spectrometer {\em (XEUS)} mission, 
currently under study to be the next ESA cornerstone mission 
for X-ray astronomy, will be capable of surveying the hard 
X-ray sky at, or even below, $10^{-14}$~ergs~cm$^{-2}$~s$^{-1}$
to energies of $40-50$~keV. Obviously, the scientific objectives 
of such a mission would be much broader than a better sampling 
of the Compton thick population. The mission would be devoted 
to probing a significant fraction of the electromagnetic 
radiation from accretion onto supermassive black holes.

\begin{acknowledgments}
The author wishes to thank Marcella Brusa, Roberto Gilli, Piero Ranalli, 
and Cristian Vignali for the extremely valuable support and informative 
discussions, Amy Barger for the opportunity to write this chapter and 
her patience, and Gianni Zamorani for a careful reading of the manuscript 
and useful comments. A special thanks to the {\tt HELLAS2XMM} team  
for the extremely pleasant collaboration.   
\end{acknowledgments}

\end{document}